\documentclass{aa}             
\usepackage{graphicx}
\usepackage{epsfig}
\usepackage{txfonts}

\def\rxj0513{RX\,J0513.9--6951}
\def\srxj0513{RXJ0513}

\begin{document}

\title{Variability in the cycle length of the supersoft source\\
\rxj0513}
\authorrunning{V. Burwitz et al.}
\author{V. Burwitz\inst{1}, 
K. Reinsch\inst{2}, 
J. Greiner\inst{1}, 
E. Meyer-Hofmeister\inst{3},
F. Meyer\inst{3},
F.M. Walter\inst{4},
R.E. Mennickent\inst{5}
}
\offprints{V. Burwitz \email {burwitz@mpe.mpg.de} }

\institute{ Max-Planck-Institut f\"ur extraterrestische Physik,
P.O. Box 1312, D-85741 Garching, Germany
\and
Institut f\"ur Astrophysik, Georg-August-Universit\"at G\"ottingen, 
Friedrich-Hund-Platz 1, D-37077 G\"ottingen, Germany
\and
Max-Planck-Institut f\"ur Astrophysik, Karl-Schwarzschild-Str.~1, D-85740 Garching, Germany
\and
Department of Physics and Astronomy, 
State University of New York at Stony Brook, Stony Brook, NY
11794-3800, USA
\and
Departamento de F\'{i}sica, Universidad de Concepci\'{o}n, Casilla 160-C,
Concepci\'{o}n, Chile  
}

\date{Received: 22.12.2006 / Accepted: 28.01.2008}


\abstract
   {The supersoft X-ray binary \rxj0513 shows cyclic
    changes between optical-low\,/\,X-ray-on states and optical-high\,/\,X-ray-off states.
    It is supposed to be accreting close to the Eddington-critical limit 
    and driven by ``accretion wind evolution''.
    }
   {We seek to derive the variations in the characteristic time scales of
   the long-term optical light curve and to determine the implications for the
   physical parameters of the system.
    }
   {We used existing and new optical monitoring observations covering a
   total time span of 14 years and compared the durations of the low and high
   states with the model calculations of Hachisu \& Kato.
    }
   {The cycle lengths and especially the durations of the optical high 
   states show a longterm modulation with variations that, according to the 
   accretion wind evolution model, would imply variations in the mass transfer 
   rate by a factor of 5 on timescales of years.
    }
   {}

 
\titlerunning {Variability in the supersoft source \rxj0513}

\maketitle{  }

\section{Introduction}
The supersoft X-ray source \rxj0513 (hereafter \srxj0513) was discovered with the
ROSAT satellite (Schaeidt, Hasinger \& Tr\"umper \cite{SchHasTru93}). This object
is the most luminous of the known supersoft X-ray binaries in the
Milky Way and the Magellanic Clouds. The source was identified with a~16.7 
magnitude emission line star (Pakull et al. \cite{Paketal93}, Cowley et al. \cite{Cowetal93}).
A few years later the monitoring capabilities of the MACHO Project (Alcock et al. \cite{Alcetal96})
revealed the unique feature which made this supersoft source a key object: the 
source has alternating high and low states of optical brightness and
appears as a supersoft object during the optical low states. The
high state lasts about 140 days typically, the low state about 40 days
with a change in brightness by 0.8 to 1.0 magnitudes within a few days.
Only the long-term optical photometry revealed that the
supersoft X-ray outbursts occurred at times of low optical
light. From this fact Southwell et al. (\cite{Souetal96}) concluded that the most likely
cause of the X-ray outburst is a contraction of the white dwarf 
atmosphere from an expanded state to a steady shell burning phase as originally suggested by 
Pakull et al. (\cite{Paketal93}). In this concept, supersoft radiation could be 
triggered by a reduced accretion rate.

Van den Heuvel et al. (\cite{vdHetal92}) showed that the ultrasoft X-ray
emission observed in supersoft sources can be explained by steady
nuclear burning of hydrogen accreted onto the surface of a white dwarf
in the mass range 0.7 to 1.2 solar masses. 
Processing hydrogen into helium at the rate of accretion requires a 
minimum mass accretion rate of $\sim 1 \times 10^{-7} M_{\sun}$/yr. 
Below this, burning is unstable and occurs in flashes. At $\sim 4 \times 10^{-7} M_{\sun}$/yr
the accretion rate approaches the Eddington critical limit.
Kato \& Hachisu (\cite{KatHac94}) have shown that optically thick wind
solutions exist which allow that steady nuclear burning on the white 
dwarf surface can continue even at super-Eddington accretion rates.

\begin{figure*}[t]
\label{fig:licu}
\epsfig{angle=0,width=18cm,file=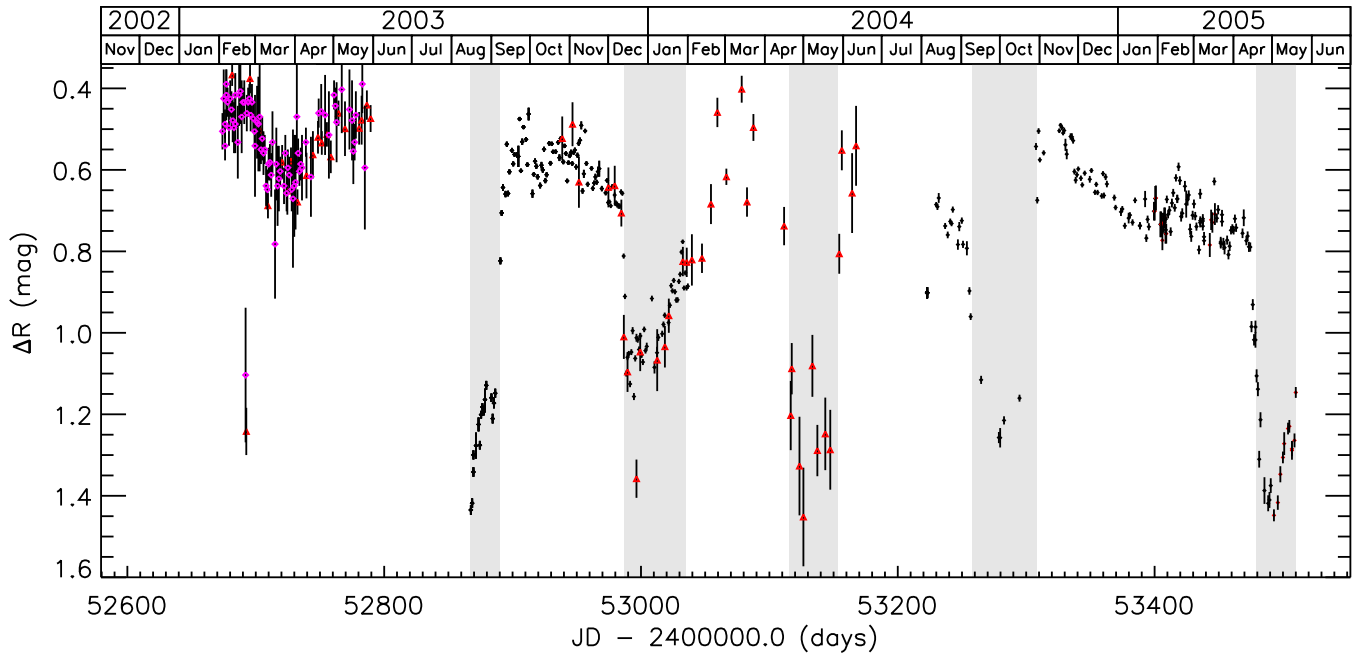,clip=}
\caption[.]{Optical monitoring:  R-band lightcurve of
            \rxj0513 obtained 
            with the 1.3-m telescope + ANDICAM and the 1.0-m telescope
            + AP7b CCD camera 
            at the Cerro Tololo Interamerican Observatory (CTIO), Chile 
            during SMARTS and Chilean observing time. 
            In addition, the photometry presented in McGowan et al. 2005 
	    (Barfold  Observatory - small triangles, CTIO 1.3-m data - small diamonds), 
            offset to match our data, is included.      
            The times of the five observed optical low states (shaded in grey) are listed in  
            Table~1.  
             
 }
\end{figure*}
Reinsch et al. (\cite{ReivTeBeu96}) demonstrated that the observed 
optical and X-ray flux variations in \srxj0513 can be quantitatively explained 
by variations in the irradiation of the accretion disk caused by
a contracting and expanding photosphere of the accreting white dwarf with
nuclear burning at its surface.
Also the lightcurve analysis of Meyer-Hofmeister \& Schandl
(\cite{MeySch96}) confirmed the importance of irradiation during the high state.
But what causes the changes between the high and low optical state,
or, between higher and lower mass accretion rates?

Reinsch et al. (\cite{Reietal00}) presented a self-maintained 
limit-cycle model which can qualitatively explain the
optical and soft X-ray variability of \srxj0513 . 
They proposed that the behavior results from periodic changes of the
accretion disk viscosity in response to changes of the irradiation by the
hot white dwarf photosphere. In this model, the mass-flow rate at the
surface of the white dwarf varies while the mass transfer from
the companion star remains constant during the cycle. 

Using their optically thick wind theory and OPAL opacities, Kato \& 
Hachisu (\cite{KatHac94}) have shown that at levels already slightly below the Eddington
luminosity an envelope around the white dwarf can no longer be static. Instead,
optically thick strong winds will onset from the white dwarf when a critical 
accretion rate of $\sim 1 \times 10^{-6} M_{\sun}$/yr is exceeded (Hachisu et 
al. \cite{HacKatNom96}). Based on this work, a new self-sustained 
transition mechanism for the cyclic behavior of \srxj0513 was developed
by Hachisu \& Kato (\cite{HacKat03a}, \cite{HacKat03b}). In this model,
excess matter above the critical accretion rate is expelled from the binary by
winds and the system evolves according to the accretion wind solution (Hachisu 
\& Kato \cite{HacKat01}). The mass-transfer rate and with it the wind mass-loss 
rate itself are periodically modulated as the strong wind interacts with the 
surface layer of the secondary star. Variations in the two rates occur with a delay 
between them due to the viscous time-scale of the accretion disk.

Both models (Reinsch et al. \cite{Reietal00}, Hachisu \& Kato \cite{HacKat03b}) 
allow predictions to be made on the variations in the temperature, the effective 
photospheric radius, and the accretion rate of the white dwarf during the X-ray 
and optical cycle of \srxj0513 which can be tested by dedicated multi-wavelength 
observations. In addition, the detailed parameter study contained in the 
Hachisu \& Kato (\cite{HacKat03b}) paper provides implications for the physical
parameters of the binary system which can be evaluated by a careful analysis of
the long-term optical light curve alone.

Two independent monitoring campaigns aimed at obtaining high-resolution X-ray
spectroscopy of \srxj0513 have been initiated to gain further insight into the 
complex nature of this supersoft source. Both campaigns use optical monitoring 
to trigger X-ray observations during an optical low state of the source. The 
first campaign resulted in XMM-Newton observations in April/ May 2004 reported
by McGowan et al. (\cite{McGetal05}). A second optical monitoring campaign has
been carried out by us to trigger X-ray observations with the {\it Chandra}
X-ray observatory, performed in December 2003 and in April/ May 2005.
First results of the monitoring and the X-ray observations were presented in 
Burwitz et al. (\cite{Buretal07}) and Reinsch et al. (\cite{ReiBurSch06}).
A detailed discussion of the X-ray observations and an analysis of the
simultaneous X-ray and optical data will be given in a separate paper (Burwitz 
et al. \cite{Buretal08}).

In this paper, we will focus on the discussion of the combined optical 
monitoring data from our own campaign and from all campaigns available in the 
literature. Altogether these data cover the optical lightcurve of \srxj0513 
over a time span of about 14 years and allow for the first time a detailed 
analysis of the varying lengths of the optical high and low states.
In Sect. 2 we give a description of the optical observations that are used. 
We compare the observations with the predictions of the Hachisu \& Kato 
(\cite{HacKat03b}) model (Sect. 3) and discuss the implications for the 
long-term cycles in the light curve on the expected changes of the mass transfer 
rate (Sect. 4).

\section{Observations}

The optical monitoring data used in this paper come from three different sources:
our own (described below), those published in McGowan et al (\cite{McGetal05}),
and the MACHO data (Cowley et al. \cite{Cowetal02}). 

Our optical data were obtained during two  
monitoring programs using the telescopes run by the SMARTS Consortium 
at Cerro Tololo, Chile.
We used the ANDICAM dual-channel photometer, mounted on the 
SMARTS/CTIO 1.3 m telescope, to obtain the CCD images
during the first monitoring program from August 2003 to January 2004 and to
obtain part of the data during the second monitoring program from August 2004 
to May 2005. Most of the data during the second campaign
were obtained with an Apogee AP7b CCD camera on the CTIO 1.0 m telescope.
In both programs B and R images were taken every 1-2 days. 
In addition V images were taken every 3-4 days during the first campaign.

Aperture photometry, using the MIDAS data analysis package, 
was used to measure the fluxes of \srxj0513 and
four comparison stars. 
Differential magnitudes in the R-filter are shown in Fig.~\ref{fig:licu}.  

The McGowan et al. (2005) data cover most of the gap between our
two monitoring programs from February to July 2004.   
From December 2002 to June 2004 they obtained
V-band and unfiltered observations from SAAO (Sutherland, South
Africa), CTIO (La Serena, Chile), and Barfold Observatory (Victoria,
Australia).  

The MACHO data described in Cowley et al. (\cite{Cowetal02}) cover 
the time span from July 1992 up to January 2000 and provide a 
close to 8 year baseline of long-term monitoring observations of \srxj0513.  

\begin{figure}
\epsfig{angle=90,width=9.0cm,file=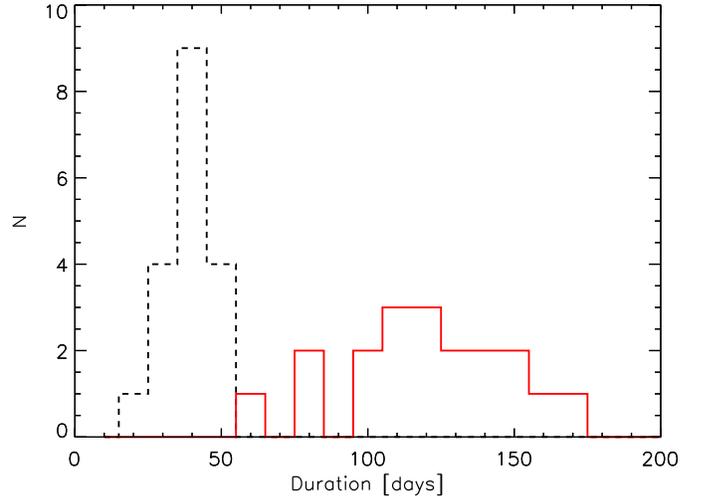} 
\caption[.]{\label{fig:dists}Histogram showing the distribution of
high 
(solid line) and low (dashed  line) state 
lengths. 
Only state lengths without upper/lower limits, taken from Table \ref{table:1}, are used in the histograms.}
\end{figure}

For our analysis of the variability in the
cycle length the time of change between high and low
states was extracted from all available observations in the same way:
we take the date when the luminosity shows the steepest increase/ decrease. 
With this method we can determine the date of state change to $\pm$1 day.
In Table \ref{table:1} we list the dates of changes between the states and their 
duration. 
No observations exist for the low state expected between JD\,2449321 and 
JD\,2449370. Therefore, only lower and upper limits can be given in
Table 1 for its onset and end, respectively. Also the changes to the low state around
JD\,2452867 and the change to the high state at JD\,2453509 are uncertain.
The distribution of both high and low state lengths is 
displayed as a histogram in Fig.~\ref{fig:dists}. 

\begin{table}
\caption{Dates of change, and duration of the optical high and low states obtained from 
the optical observations presented in Section~2. The distribution of the state lengths is 
displayed as a graph in Fig.~\ref{fig:dists}.}             
\label{table:1}      
\begin {center}          
\begin{tabular}{c c c c}     
\hline\hline                      

change down    & change up to	    & duration of   & duration of  \\ 
to low state   & high state	    & low state     & high state$^+$ \\  
JD-2400000    & JD-2400000	    & (days)	    & (days)	   \\
\hline 
    ---     &    48831  &  ---  &     \hspace*{2mm}78 \\
   48909   &    48937  &     28 &     \hspace*{2mm}63 \\
   49000   &    49038  &     38 &    119 \\
   49157   &    49198  &     41 & $>$123\hspace*{2mm} \\
$>$49321\hspace*{2mm}  & $<$49370\hspace*{2mm}  &  $<$49\hspace*{2mm} & $>$149\hspace*{2mm} \\
   49519   &    49551  &     32 &    141 \\
   49692   &    49725  &     33 &    137 \\
   49862   &    49901  &     39 &    108 \\
   50009   &    50040  &     31 &    110 \\
   50150   &    50190  &     40 &    116 \\
   50306   &    50344  &     38 &    156 \\
   50500   &    50537  &     37 &    154 \\
   50691   &    50728  &     37 &    153 \\
   50881   &    50928  &     47 &    130 \\ 
   51058   &    51080  &     22 &    118 \\
   51198   &    51243  &     45 &    104 \\
   51347   &    51391  &     44 &    129 \\
   51520   & $>$51546\hspace*{2mm} &  
                          $>$26\hspace*{2mm}   &   ---	 \\
   ...   &    ...  &    ...    &  ...	 \\
$<$52867\hspace*{2mm} &    
                52890  &  $>$23\hspace*{2mm}  & 
                                     100 \\
   52990   &    53037  &     47 &     \hspace*{2mm}78 \\
   53115   &    53153  &     38 &    105 \\
   53258   &    53308  &     50 &    171 \\
   53479   & $>$53509\hspace*{2mm}   &     $>$30\hspace*{2mm}   &    ---    \\
\hline                  

\end{tabular}
\end {center}
$^+$ next high state\\

\end{table}

\section{Comparison with the model of Hachisu \& Kato}

\subsection{The sequence of states of nuclear burning in the model} 

In the model of Hachisu \& Kato (\cite{HacKat03a}, \cite{HacKat03b})
diverse physical processes have to be included to reproduce
the cycles present in  \srxj0513.
The basic input is the detailed computational results on 
the nuclear burning on the white dwarf surface (Hachisu \& Kato \cite{HacKat03b}),  
which are based on the analysis of optically thick winds in nova outbursts 
(Kato \& Hachisu \cite{KatHac94}). For the extended
envelope a special surface boundary condition was chosen,
combining the structure of a static envelope with maximal (Eddington) 
luminosity and a wind solution in which the sonic point is close to
the photosphere. Using the results for  given white dwarf mass and chemical
abundances several physical quantities can be evaluated as 
a function of the envelope mass:
the rate of nuclear burning, the white dwarf radius, the photospheric
temperature, the luminosity, and the wind loss rate. 
These relations allow us to determine the change of envelope mass with
time. Hachisu \& Kato (\cite{HacKat03a}, \cite{HacKat03b}) assume that
as a consequence of the strong wind the surface layer of the main sequence 
companion is stripped off. The mass transfer to the disk around
the white dwarf is reduced by the stripping rate and stops if the latter
becomes larger than the original mass-transfer rate. After a viscous 
timescale of the disk also the accretion onto the white dwarf stops, 
the wind ceases, the mass-transfer is resumed, and a new cycle begins.

\subsection {The optical high/ X-ray low state}
An important contribution to the luminosity during the optical high state 
is expected from a large and irradiated disk surface, 
which leads to a quick rise of the V luminosity. 
Hachisu \& Kato (\cite{HacKat03a}, \cite{HacKat03b}) assumed 
that the disk size increases to three times the Roche lobe radius 
of the white dwarf (actually only a large radiating area is needed to
explain the luminosity increase, the geometry is not constrained). 
This might appear as an ad hoc assumption, but such a
contribution was found as appropriate for the fitting of the lightcurve 
of CI Aql in the outburst decline (Hachisu \& Kato \cite{HacKat03a}).

The duration of the optical high/ X-ray low state is expected to depend
mainly on the accretion rate. Monitoring observations have shown that this
duration varies by more than a factor of 2 and accordingly large changes in the
accretion rate from the secondary are expected to occur (see Section 4).

\subsection {The optical low/ X-ray high state}
At the end of the wind phase, the system enters an optical low/ X-ray high 
state. During this phase, supersoft radiation from the nuclear-burning white 
dwarf surface can emerge while optical emission is reduced by about 1 magnitude as
the accretion disk has shrunk to its original size. While accretion from the
secondary resumes as soon as the wind stops, it takes the viscous timescale 
of the accretion disk for the accreted material to reach the white dwarf 
surface. Therefore, this timescale largely determines the duration of the
optical low/ X-ray high state.
The radius of the white dwarf is a function of the
amount of matter in the hydrogen layer on the white dwarf surface
(Kato \cite{Kat85}, Kato \& Hachisu \cite{KatHac94}). The analysis of
the X-ray data shows that the radius and temperature of the white dwarf
roughly agree with the predictions of the Hachisu \& Kato
(\cite{HacKat03b}) model (see Burwitz et al. \cite{Buretal07}). 

\begin{figure}

\hspace*{0.7cm}\includegraphics[width=6.0cm,angle=90]{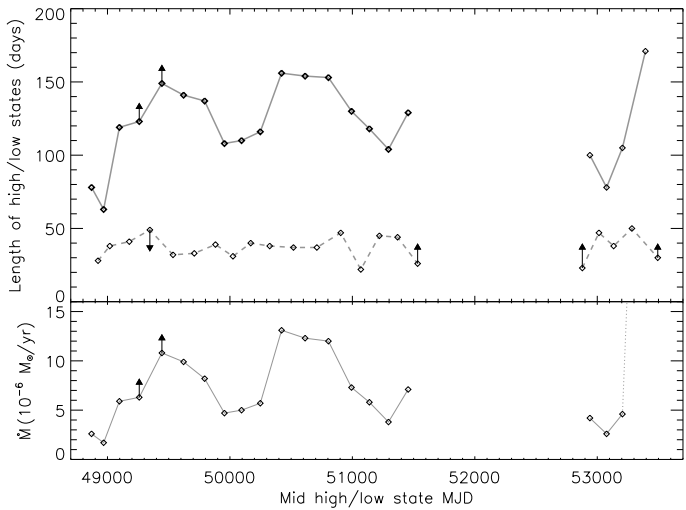}
\vspace*{0.5cm} 
\caption[.] {\label{fig:durst}Upper panel: duration of the observed optical high 
(solid line) and low (dashed line) states in the years 1992 to
2005. Lower panel: mass transfer rates inferred from the results of 
the Hachisu \& Kato model for the dependence of cycle lengths on mass 
transfer rates. State lengths for which we only have upper/lower limits in Table 1
are indicated by arrows.}
\end{figure}

\subsection{What causes the cyclic behavior?}
Essential in the model of Hachisu \& Kato (\cite{HacKat03a}, \cite{HacKat03b})
is the assumption that the mass accretion onto the white dwarf is
intermittent. This assumption causes a cyclic behavior. If,
otherwise,  the mass transfer could adjust to a certain amount of the wind 
mass loss the source could remain at this state. The same question can be 
raised in connection with the model of Reinsch et al. (\cite{Reietal00}). Could
the disk exist in a stable state with a certain amount of irradiation 
and with a mass flow rate according to the transfer rate?

The question arises whether a hysteresis in the relation between
optical radiation and amount of matter in the hydrogen layer could
cause a cyclic behavior. Can the luminosity be different for
different atmospheric structures, in one case for an extended
photosphere (determined by wind loss), 
and in another case for a star not blown up (with the same amount of
hydrogen at both stages)? The results for nova outbursts (Kato \&
Hachisu 1994) seem to yield a unique relation. The only free parameter
taken is the value for the optical depth at the photosphere (as
mentioned above), 
chosen in a way that the structure of the maximum static solution and 
that of the minimum wind solution are almost identical. 

We performed a stability analysis for the equilibrium flow in the 
Hachisu \& Kato model. 
The change of the envelope mass ${d\Delta M}/{dt}$ depends on the rate of 
mass burning on the white dwarf surface, the wind mass loss in Kato's 
solution and the accretion rate on the white dwarf surface. Wind mass loss
and burning rate are a function of the envelope mass and increase with the 
envelope mass, while mass transfer decreases with the increasing wind 
mass loss. This provides 
for a stationary solution. The finite diffusion time through the accretion 
disk introduces a time-lag between mass transfer from the secondary star and 
the accretion on the white dwarf surface. (For the detailed analysis
see Appendix A.)  

The result is that the
stationary solution is stable: The mass accretion onto the white dwarf
always can adjust to the wind from the photosphere, small deviations
from the stationary state decay exponentially. 
In the model of Hachisu \& Kato (\cite{HacKat03a}, \cite{HacKat03b}) the cyclic behavior 
is introduced by an assumed hysteresis in the interaction between 
wind and mass transfer from the secondary star. 

\begin{figure}

\hspace*{0.7cm}\includegraphics[width=6.0cm,angle=90]{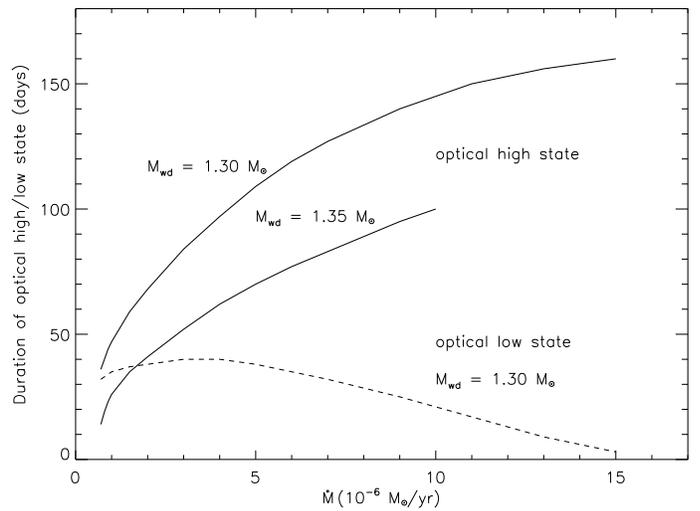}
\vspace*{0.5cm}  
\caption[.]{\label{fig:depcl}Relation between mass transfer rate and duration of states
(results from the model of Hachisu \& Kato 2003b). Solid lines: duration of optical high-states for 
different white dwarf masses. Dashed line: duration of optical low states.}
\end{figure}

\section{Interpretation of the observed long-term evolution of cycle
length} 
In general the duration of the optical high state is determined by
the burning of the hydrogen in the envelope during the wind phase. 
The higher the rate of mass accretion onto the white dwarf during this
time, that is, the higher the overflow rate from the companion star
during the pre-wind phase, the longer the high state lasts. 
The length of the high states varies from about 63 days to 171 days in the recent cycle in
2005. 
In Fig.~\ref{fig:durst} 
(upper panel) we show the observed duration of
optical high and low states as listed in Table~\ref{table:1}.
Hachisu and Kato (\cite{HacKat03b}) used their model to calculate lightcurves for 
various mass transfer rates and determined the cycle length. 
In Fig.~\ref{fig:depcl} we show their results for the dependence of the cycle 
length on the mass transfer rate (we take the results for the  assumed
viscous timescale of the disk $t_{\rm vis}$\,=\,20.5\,days which agrees best  with the observations). 

\begin{figure*}
\epsfig{angle=90,width=18.0cm,file=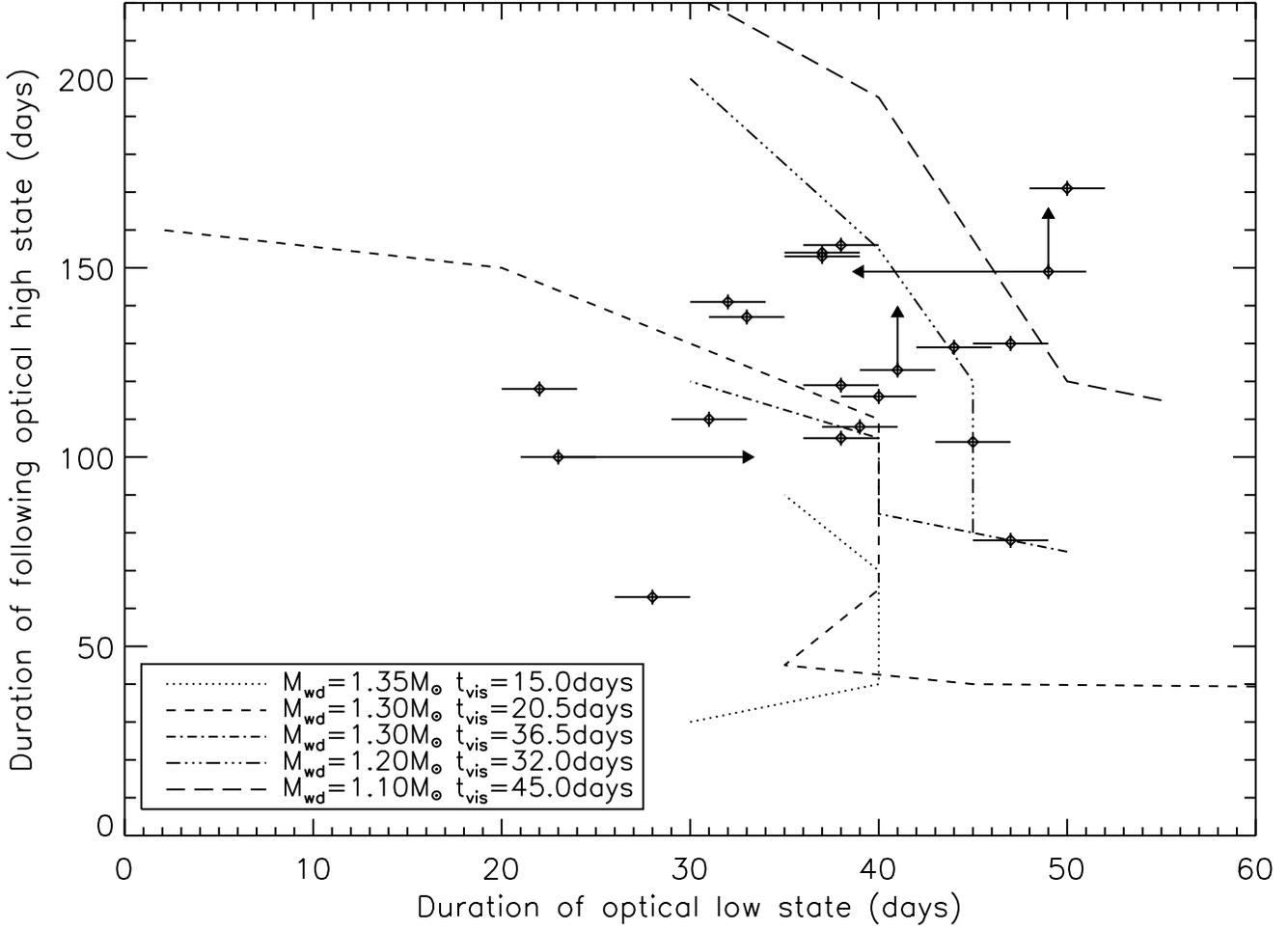,clip=} 
\caption[.]{\label{fig:durhilo} Observed duration of the optical low states
compared to the duration of the immediately following optical high states
(data points). The continuous lines illustrate the expected dependance between 
both ``durations'' as a function of mass-transfer rate as tabulated by Hachisu \& Kato 
2003 for various assumptions on the parameters white dwarf mass and viscous 
timescale. State lengths for which we only have upper/lower limits taken from Table 1
are indicated by arrows.
}
\end{figure*}

Taking the relation between the length of high
states and the mass accretion rate from Fig.~\ref{fig:depcl}
we can interpret the observed duration of 
the high states from Table \ref{table:1} as a measure for the 
mass transfer rate (Fig.~\ref{fig:durst}, lower panel). 
As can be seen the mass transfer rate is expected to vary by a factor 
up to 5 and a long-term modulation of the mass transfer rate appears
to be indicated. In the framework of the model, we note particularly that 
the recently observed long cycle would be related to an episode of higher mass accretion 
rate.

According to the numerical calculations of Hachisu \& Kato 
(\cite{HacKat03b}) the duration of the optical low state depends also 
on the mass accretion rate, but with a higher rate leading to a 
shorter duration. 
In their model, mass transfer into the potential well of the white dwarf 
occurs only during the optical low states. Therefore, the mass transfer rate 
during these episodes is expected to be the relevant parameter which determines 
the duration of the low state itself as well as the duration of the subsequent 
optical high state. In Fig.~\ref{fig:durhilo} we have compared our measurements
of times and show expected relations between them according to the parameter 
studies of Hachisu \& Kato (\cite{HacKat03b}). From the available data of all
observed cycles there is no clear correlation in the sense that a relatively 
short optical low state is followed by a long optical high state and vice 
versa. In this respect the observations differ from the model predictions. 

The origin of the possible long-term changes of the mass overflow rate 
is unclear. Stellar activity might be a potential mechanism. Stellar
pulsations are less likely relevant. Hutchings et al. (\cite{Hutetal02}) argue, 
in their work on ultraviolet spectroscopy of \srxj0513, that the donor
might be a more massive star. But long pulsation periods, of years, are known 
only for AGB stars.

\section{Conclusions}
The observations of \srxj0513 document long-term changes of the
duration of the optical-high\,/\,X-ray-off and the optical-low\,/\,X-ray-on
states. A model for the cyclic changes was presented by 
Hachisu \& Kato  (\cite{HacKat03a}, \cite{HacKat03b}). In this paper, we
have discussed the assumptions made in their model and compared the model 
predictions with observational results from the long-term optical monitoring 
of \srxj0513{}.

An essential question is the cause of the limit cycle behavior. Our
stability analysis (see Appendix A) shows that a stable 
stationary solution exists in which the mass accretion onto the white 
dwarf always can adjust to the wind from the photosphere. This analysis does
not support the model assumption of Hachisu \& Kato  (\cite{HacKat03a}, 
\cite{HacKat03b}) that the wind from the white dwarf stops the mass overflow
from the companion and thereby leads to the limit cycle behavior. We conclude
that an as yet unspecified hysteresis seems to be required to naturally explain 
the cyclic behavior of \srxj0513{}.

The numerical calculations by Hachisu \& Kato (\cite{HacKat03a}, 
\cite{HacKat03b}) for different mass overflow rates from the companion star 
provide a relation for the dependence of the cycle length on the mass transfer 
rate. We have used this relation to derive mass transfer rates from the lengths 
of the optical high states. In this picture, the observed variability of the 
cycle length would indicate that the mass overflow rate from the companion star
varies by a factor of about 5 on a timescale of few years. The recent very 
long optical high state leads to the question whether \rxj0513 is in a 
transition from a phase of lower mass transfer in the past to higher rates and,
maybe, even approaches an uninterrupted high state. But, looking back
to the past, observations 100 years ago (Leavitt \cite{Lea1908}) show 
already variations in the luminosity at that time.

\begin{acknowledgements}

R.E. Mennickent acknowledges the Grant Fondecyt 1030707.  
The State University New York (SUNY) Stony Brook membership
in the SMARTS consortium is made possible
by generous support from the Vice President for Research, the Provost, and the
Dean of Arts \& Sciences.

We also would like to thank the anonymous referee for valuable comments and
suggestions which helped to improve the structure of the paper.

\end{acknowledgements}

\begin{appendix}

\section{Stability analysis}
We consider the time evolution of the mass $\Delta M$ in the white dwarf envelope fed 
by mass accretion $\dot M_{\rm{acc}}$, but consumed by nuclear burning $\dot M_{\rm{b}}$ 
and by wind loss $\dot M_{\rm{w}}$,
\begin {equation} 
\frac{d}{dt} \Delta M = -\dot M_{\rm{b}} -\dot M_{\rm{w}}+\dot M_{\rm{acc}}.
\end {equation} 
Mass accretion $\dot M_{\rm{acc}}$ occurs via an accretion disk fed by mass 
transfer $\dot M_{\rm{transf}}$ from the secondary star. 
With $\dot M_{\rm{trans}}$ starting at some time $t=0$ (arbitrary), one has
\begin {equation}
\dot M_{\rm{acc}}(t) =\int\limits_{0}^{t}\dot M_{\rm{transf}}(t-s) e^{-s/\tau}\frac{ds}{\tau} 
\end {equation}
with diffusion time $\tau$ of the disk.

The model assumptions (see Hachisu \& Kato (2003a, 2003b) are: 
$\dot M_{\rm{w}}(\Delta M)$ is a monotonically rising
function of $\Delta M$. $\dot M_{\rm{transf}}(\dot M_{\rm{w}})$ is a 
monotonically
decreasing function of $\dot M_{\rm{w}}$ (through interference of the
wind with the mass transfer), such that mass transfer is shut off at
some critical wind loss rate 
$(\dot M_{\rm{w}})_{\rm{c}}$. For small $\dot
M_{\rm{w}}$, mass transfer  $\dot M_{\rm{transf}}$ is assumed larger than 
$\dot M_{\rm {w}}$. We further neglect a weak dependence of 
$\dot M_{\rm {b}}$ on $\Delta M$. 

In steady state, $\dot M_{\rm{acc}}$= $\dot M_{\rm{transf}}$, and Eq.\,A1
becomes 
\begin {equation} 
0 = -\dot M_{\rm{b}} -\dot M_{\rm{w}}(\Delta M)+\dot
M_{\rm{transf}}(\dot M_{\rm{w}}).
\end {equation} 

The right hand side monotonically decreases with increasing $\Delta
M$, is positive for small  $\Delta M$ and negative for $\Delta M =
\Delta M_{\rm {c}}$, the value at which the wind loss rate reaches the
value $(\dot M_{\rm{w}})_{\rm {c}}$ where mass transfer is shut
off. In between those exists one value $(\Delta M)_{\rm {s}}$ for which
Eq.\,A3 is fulfilled and steady state holds.

This steady state is stable. We linearize Eqs.\,A1 and A2 in deviations
 $\delta \Delta M$ and $\delta \dot M$ from equilibrium, and obtain
\begin {equation}
\frac{d}{dt} \delta \Delta M=-\frac{1}{\tau_{\rm{w}}}\left[\delta
\Delta M
+\alpha\int_{0}^{t}\delta \Delta M (t-s) e^{-s/\tau} \frac{ds}{\tau}\right]. 
\end {equation} 

Here we have used the functional dependence of $\dot M_{\rm{w}}$ on
$\Delta M$ and $\dot M_{\rm{transf}}$ on $\dot M_{\rm{w}}$, and defined
the wind loss time scale $\tau_{\rm{w}}$ and the interaction strength  $\alpha$
between wind and mass transfer as 

\begin {equation}
\frac{d\dot M_{\rm{w}}}{d\Delta M}=\frac{1}{\tau_{\rm{w}}}, 
\hspace {0.4cm}
\frac{d\dot M_{\rm{transf}}}{d\dot M_{\rm{w}}}=-\alpha.
\end {equation}

Laplace transformation (Abramowicz \& Stegun \cite{AbrSte64}) of Eq.\,A4 gives
$F(p)=\int_{0}^{\infty}e^{-pt}\delta \Delta M (t) dt$ as a rational
function of $p$ with two poles in the complex plane $p$ with negative
real parts. Inverse Laplace transformation reduces to evaluation of
the residuals at these poles and gives the solution as a damped
oscillator, 
\begin {equation}
\delta \Delta M (t)=e^{-t/\tau_*} (cos\, \omega _* t + b sin \,\omega_* t)
\: \delta \Delta M (0),
\end {equation}
$\tau_*$, $\omega _*$, and $b$ are combinations of $\tau$, $\tau_{\rm{w}}$ and $\alpha$,

\begin{eqnarray}
\tau_*&=&\frac{2\tau\tau_{\rm{w}}}{\tau+\tau_{\rm{w}}}\\
\omega_*&=&\frac{\sqrt{4\alpha \tau\tau_{\rm{w}}-(\tau_{\rm{w}}-\tau)^2}}
{2\tau\tau_{\rm{w}}},\\
b&=&\frac{\tau_{\rm{w}} -(1+2\alpha)\tau}
{\sqrt{4\alpha\tau\tau_{\rm{w}}-(\tau_{\rm{w}}-\tau)^2)}}.
\end{eqnarray}

If the expression under the square root becomes negative, Eq.\,A6
becomes that of an overdamped oscillator with two exponentially
decaying terms.

We conclude that the model does not lead to sustained cycles between
states of high and low mass transfer by time delay in the disk
accretion process alone, and that a well documented
hysteresis either in the relation $\dot M_{\rm{transf}} 
(\dot M_{\rm{w}})$, or more simply, in the dependence $\dot M_{\rm{w}}(\Delta
M)$ is required to naturally explain the cyclic behavior of the
supersoft source \rxj0513.

\end {appendix}
\end{document}